\documentclass[]{aipproc}

\layoutstyle{8x11single}

\begin{document}

\title{Flow and non-flow correlations from four-particle multiplets in STAR}

\classification{25.75-q, 25.75.Ld}
\keywords{Relativistic heavy-ion collisions; Elliptic flow; cumulant}

\author{A. H. Tang for the STAR collaboration }{
  address={Department of Physics, Kent State University, Kent, OH 44242},
  email={aihong@cnr.physics.kent.edu}
}

\copyrightyear  {2001}

\begin{abstract}
 Elliptic flow results are presented for Au + Au collisions at $\sqrt{s_{NN}} = 130$  
 GeV in RHIC.  This signal is investigated as a function of transverse momentum,  
 rapidity and centrality.  Results from four-particle correlation analysis, which  
 can filter out contributions to the flow signal from correlations unrelated to the  
 event reaction plane (``non-flow''), are presented and compared to the conventional  
 method, in which non-flow effects are treated as part of the systematic uncertainty.   
 
\end{abstract}

\date{\today}

\maketitle

\section{Introduction}

Flow describes the azimuthal momentum distribution with respect to 
the reaction plane of particle emission from non-central heavy-ion 
collisions \cite{Reis97,Herr99}.  The initial spatial deformation due 
to geometry and the pressure developed early in the collision causes 
azimuthal momentum-space anisotropy.  The measurement of flow can help 
provide insight into the evolution of this early stage of a relativistic 
heavy-ion collision.  Elliptic flow is characterized by the second harmonic 
coefficient $v_2$ of an azimuthal Fourier decomposition of the momentum 
distribution \cite{Olli92, Volo96, Posk98}, and has been observed and 
extensively studied in Au + Au collisions from subrelativistic energies 
on up to RHIC.  At RHIC energies, elliptic flow is inferred to be a 
relative enhancement of emission {\it in} the plane of the reaction, 
and provides information about the early-time thermalization achieved in 
the collisions \cite{STAR01}.  

Sideward and elliptic flow are widely studied phenomena with 
a well-understood relationship to the event reaction plane 
\cite{Reis97,Herr99}.  Generally speaking, large values of 
flow observables are considered signatures of hydrodynamic 
behavior, while smaller flow signals can have alternative 
explanations.  Furthermore, there are several possible sources 
of azimuthal correlations which are unrelated to the reaction 
plane, examples include correlations caused by resonance decays or 
dijets, by HBT or Coulomb effects, by momentum conservation, etc.  
In the present type of study, it is not necessary to distinguish between 
the various possible effects in this overall category, and their combined 
effect is known as ``non-flow'' correlations.  

Conventional flow analyses are equivalent to averaging over correlation 
observables constructed from pairs of particles, and there is no requirement 
for each event to contribute more than one pair.  When such analyses are 
applied to relativistic nuclear collisions where particle multiplicities can 
be as high as a few thousand, the possible new information contained 
in higher multiplets remains untapped.  A previous study of high-order flow 
effects focused on measuring the extent to which all fragments contribute to 
the observed flow signal \cite{Jian92}.  Given that flow analyses based on 
pair correlations are sensitive to both flow and non-flow effects, the 
present work investigates correlation observables constructed from particle 
quadruplets, and it is assumed that non-flow effects contribute at a 
negligible level to the quadruplet correlation.

This study presents STAR (Solenoidal Tracker At RHIC) data from Au + Au 
running at $\sqrt{s_{NN}} = 130$ GeV during summer 2000.  Details of the 
detector in its year-one configuration can be found elsewhere 
\cite{STAR99,STAR01}.  The present analysis is based on 
120k events corresponding to a minimum bias trigger.  
Events with a primary vertex beyond 1 cm radially from the center of the 
beam or 75 cm longitudinally from the center of the Time Projection Chamber 
(TPC) were excluded.  Within the selected events, tracks were included if 
all five of the following conditions were satisfied: 
they passed within 2 cm of the primary vertex, they had at least 15 
space points in the TPC, the ratio of the number of space points to the 
expected maximum number of space points was above 0.52, pseudorapidity 
$|\eta| < 1.3$, and $0.1 < p_t < 2.0$ GeV$/c$.  The above cuts are 
essentially the same as used in the previous STAR studies of elliptic flow 
\cite{STAR01,STAR01b}.

\section{Two- and four-particle correlation methods}

The conventional flow analysis methods \cite{Dani88,Olli92,Posk98} are 
based on measurement of the correlation between particle pairs, and in past 
analyses where the non-flow contribution was of concern, it was estimated 
independently.  In the first study of elliptic flow in STAR 
\cite{STAR01}, the non-flow effect from jets and resonances was 
estimated by assuming that they contribute to the second harmonic at 
the same level as to the first harmonic, and this established an upper limit 
on the non-flow contribution to the reported $v_2$ signal.  This limit played 
a role in determining the systematic error on the published measurements.   

The cumulant \cite{Biya81,Libo89,Egge93,Olli01,borghiniCumuG} and generating function approach 
offers a formal and convenient way to study flow and non-flow contributions 
systematically.  In this method \cite{Olli01}, the cumulant to order four is 
defined by 

\begin{equation}
\langle\langle e ^{in ( \phi_1 + \phi_2 - \phi_3 - \phi_4 ) } \rangle\rangle 
\equiv  
\langle e ^{in (\phi_1 + \phi_2 - \phi_3 - \phi_4 ) } \rangle  - 
\langle e ^{in (\phi_1 - \phi_3) } \rangle \langle e ^{in (\phi_2 - \phi_4) } \rangle - 
\langle e ^{in (\phi_1 - \phi_4) } \rangle \langle e ^{in (\phi_2 - \phi_3) } \rangle
\,,
\end{equation}
where the double angle bracket notation represents the cumulant expression 
shown explicitly on the right-hand side, $\phi$ is the measured azimuthal angle for 
an individual particle, and the $n$ is the harmonic whose coefficient is being 
studied.  The cumulant 
$\langle\langle e ^{in (\phi_1 + \phi_2 - \phi_3 - \phi_4) }\rangle\rangle$
involves only pure four-particle correlations since the two-particle 
correlations among the quadruplets have been explicitly subtracted away.  

In the presence of flow,  the cumulant \cite{Olli01} becomes

\begin{equation}
\langle\langle e ^ {in (\phi_1 + \phi_2 - \phi_3 - \phi_4) } \rangle\rangle 
= - v^4_n + O ( \frac{1}{M^3} + \frac{v^2_{2n}}{M^2} )
\,,
\end{equation}
where $M$ is the multiplicity of the events.  
The cumulant to higher orders and the corresponding generalization of the above 
can also be determined.  Likewise, the cumulant of order two reduces to the 
equivalent of a pair correlation analysis of the conventional type.  Statistical 
uncertainties associated with a cumulant analysis increase with increasing order.  
We find that statistics from STAR year-one data are adequate for this preliminary 
study of the 4th-order cumulant, but it is not yet feasible to investigate orders 
higher than four using available STAR data.

An alternative approach to distinguishing flow and non-flow contributions to 
$v_2$ is to partition each event into four subevents randomly.  

\begin{equation}
\langle Q_1 Q^*_2 Q_3 Q^*_4 \rangle - 2 {\langle Q_1 Q^*_2 \rangle}^2 
= 
\langle v^4_n \rangle -2{ \langle v^2_n \rangle }^2
\,,
\end{equation}
where the vector $Q_j$ is the summation of $e^{in\phi}$ for all tracks in  
subevent $j$ divided by the multiplicity for that subevent.  
This method can be implemented to estimate the non-flow effect, and yields 
results that are consistent with those from the cumulant approach. However the 
statistical error from four subevent method is relatively large since it uses
only a fraction of all possible quadruplets.

\begin{figure}[h]
  \resizebox{30pc}{!}{\includegraphics{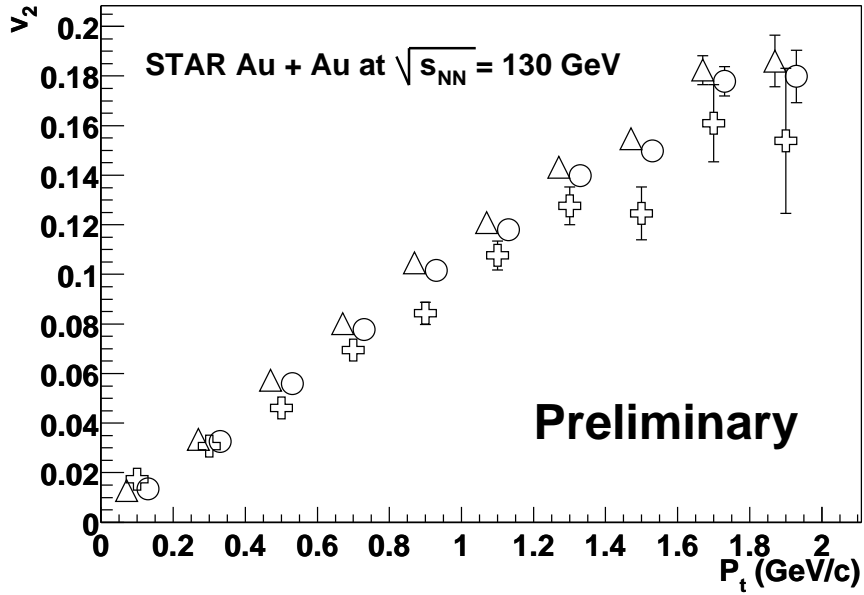}}
\caption{ $v_2$ as a function of transverse momentum for events with charged 
particle multiplicity near the middle of the observed range (see text). 
The circle, triangle and cross represent $v_2$ from the conventional method, 
from the 2nd-order cumulant method, and from the 4th-order cumulant method, 
respectively.}
\end{figure}

\begin{figure}[h]
  \resizebox{30pc}{!}{\includegraphics{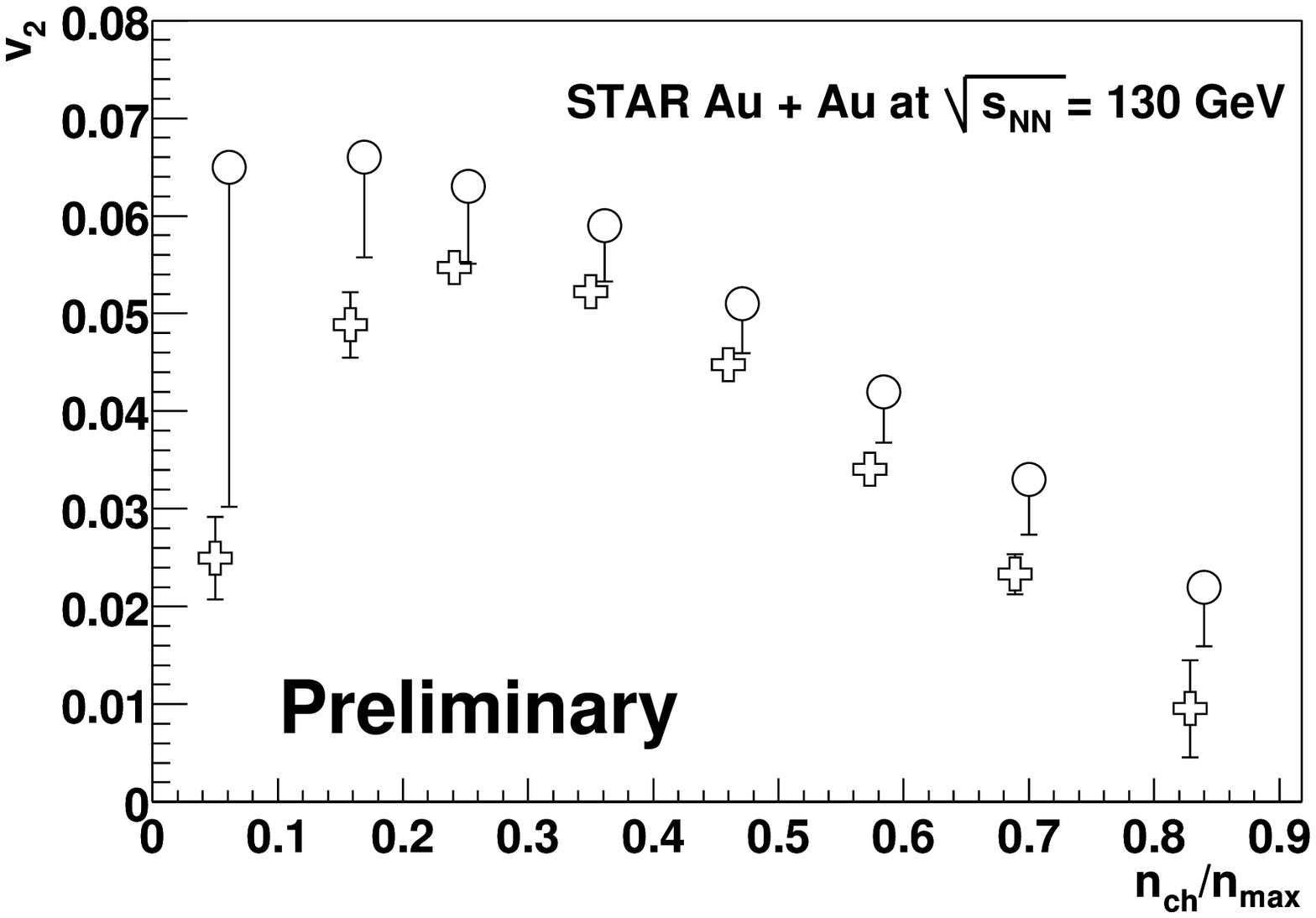}}
\caption{ $v_2$ as a function of centrality, where centrality is characterized 
by charged particle multiplicity $n_{\rm ch}$ divided by the maximum observed 
charged particle multiplicity, $n_{\rm max}$.  The circle and cross 
represent conventional $v_2$ and the 4th-order cumulant $v_2$, respectively.}
\end{figure}

Figure 1 shows $v_2$ as a function of $p_t$ for events whose centrality lies
between 0.26 and 0.34 on the scale of charged particle multiplicity 
$n_{\rm ch}$ normalized by maximum observed charged particle multiplicity, 
$n_{\rm max}$.  Figure 1 demonstrates that the two pair correlation methods 
--- conventional $v_2$ and the 2nd-order cumulant $v_2$ are consistent with 
each other, as required.  The 4th-order cumulant $v_2$ is systematically lower 
than the other two calculations, verifying that non-flow effects contribute 
to the conventional $v_2$ analysis \cite{Posk98,Mai,borghiniNoflow}.  When this comparison is repeated  
for central events and again for peripheral events, the statistical error on 
the 4th-order cumulant $v_2$ becomes bigger, but the same pattern can be 
observed.

The observed $v_2$ as a function of pseudorapidity (not shown) is almost 
flat within $|\eta | < 1.3$ for the same three methods, and again, 
the quadruplet calculation lies below the two $v_2$ observables based 
on pair correlations.

The integrated $v_2$ as a function of centrality is shown in Fig.~2.  The 
circles show the conventional $v_2$, with the reported systematic uncertainty 
\cite{QM2001Raimond} represented by the asymmetric error bars.  The statistical 
error for the conventional method is smaller than the symbol size. The crosses
show the 4th-order cumulant $v_2$, which is consistent within statistical
uncertainties with the expectation that $v_2$ corrected for non-flow effects
should lie at or within the systematic uncertainties reported for the previous
conventional $v_2$ measurements \cite{STAR01}.  It is clear that non-flow
effects are present at all centralities, and their size is largest for
peripheral collisions, as expected \cite{Olli01}. 

In order to test the reliability of the cumulant results, various simulated 
events have been processed via the same analysis procedure as the data.  The 
results for simulated events are shown in Figs.~4 and 5.  Nine data sets were 
produced, each having 10k simulated events of constant multiplicity $n_{ch} = 
500$, with $v_2 = 0.10$.  Then, a simple non-flow effect consisting of embedded 
back-to-back track pairs was introduced at various levels ranging up to 80 
embedded pairs per simulated event.  Figure 4 illustrates that the 4th-order 
cumulant $v_2$ always recovers the input 10\% $v_2$, while the $v_2$ from 
the pair correlation analysis methods can only recover the correct input if 
non-flow pairs are not embedded.  Figure 5 reports results of a similar test, 
except that here, the number of embedded pairs was constant at 50 per event, 
while the imposed input level of elliptic flow was varied.  Again, it is seen 
that the reconstructed $v_2$ from the 4th-order cumulant analysis agrees with 
the input elliptic flow, while the other two methods are biased towards 
overestimating the true $v_2$ flow signal. 

\begin{figure}[h]
  \resizebox{30pc}{!}{\includegraphics{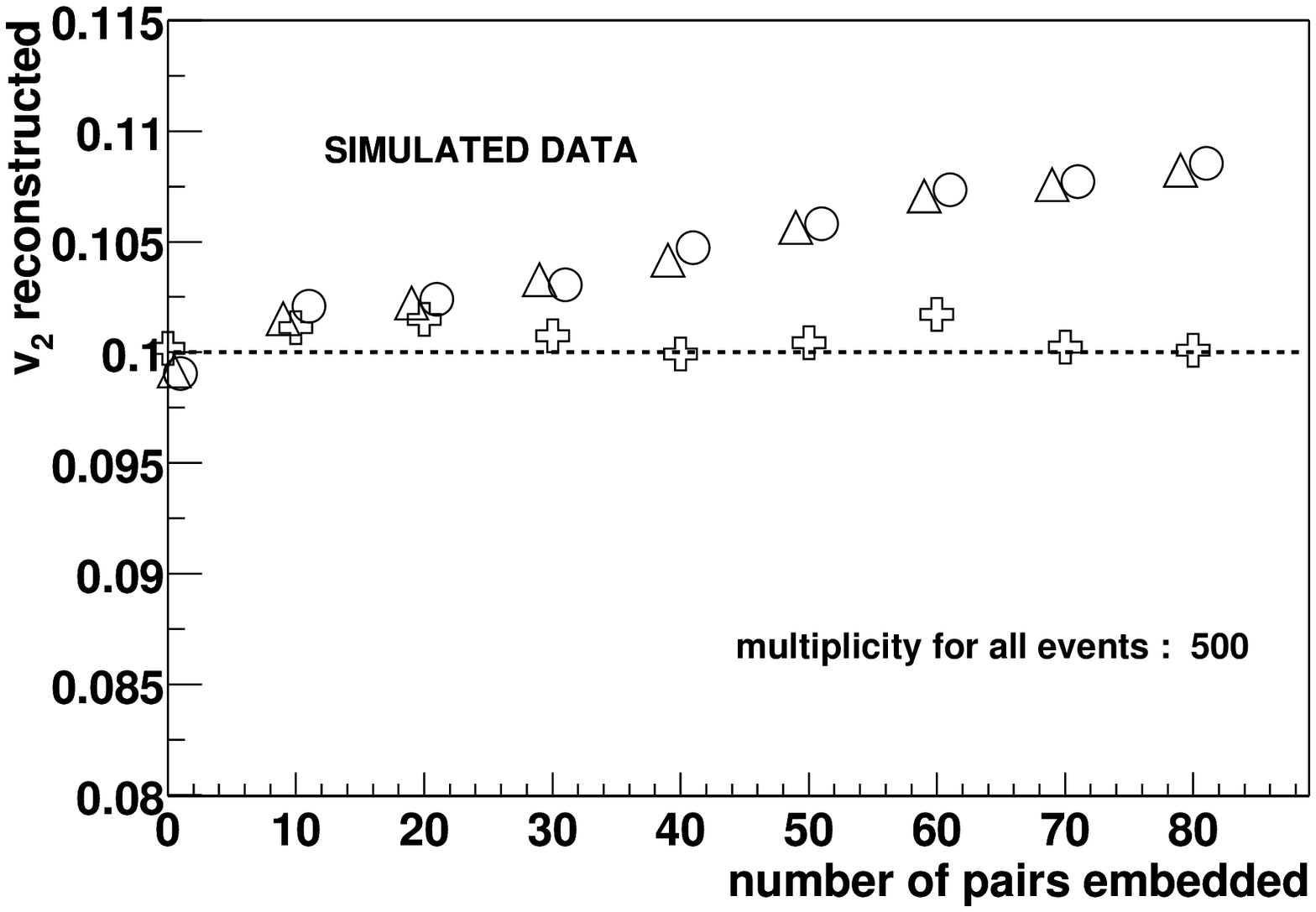}}
\caption{ Reconstructed $v_2$ for simulated events as a function of number of 
embedded back-to-back track pairs.  The horizontal dashed line marks the level 
of the true elliptic flow $v_2 = 0.10$, as imposed on the simulated events.  
The circles, triangles and crosses represent $v_2$ from the conventional 
method, from the 2nd-order cumulant method, and from the 4th-order cumulant 
method, respectively. The statistical error is smaller than the symbol size. }
\end{figure}

\begin{figure}[h]
  \resizebox{30pc}{!}{\includegraphics{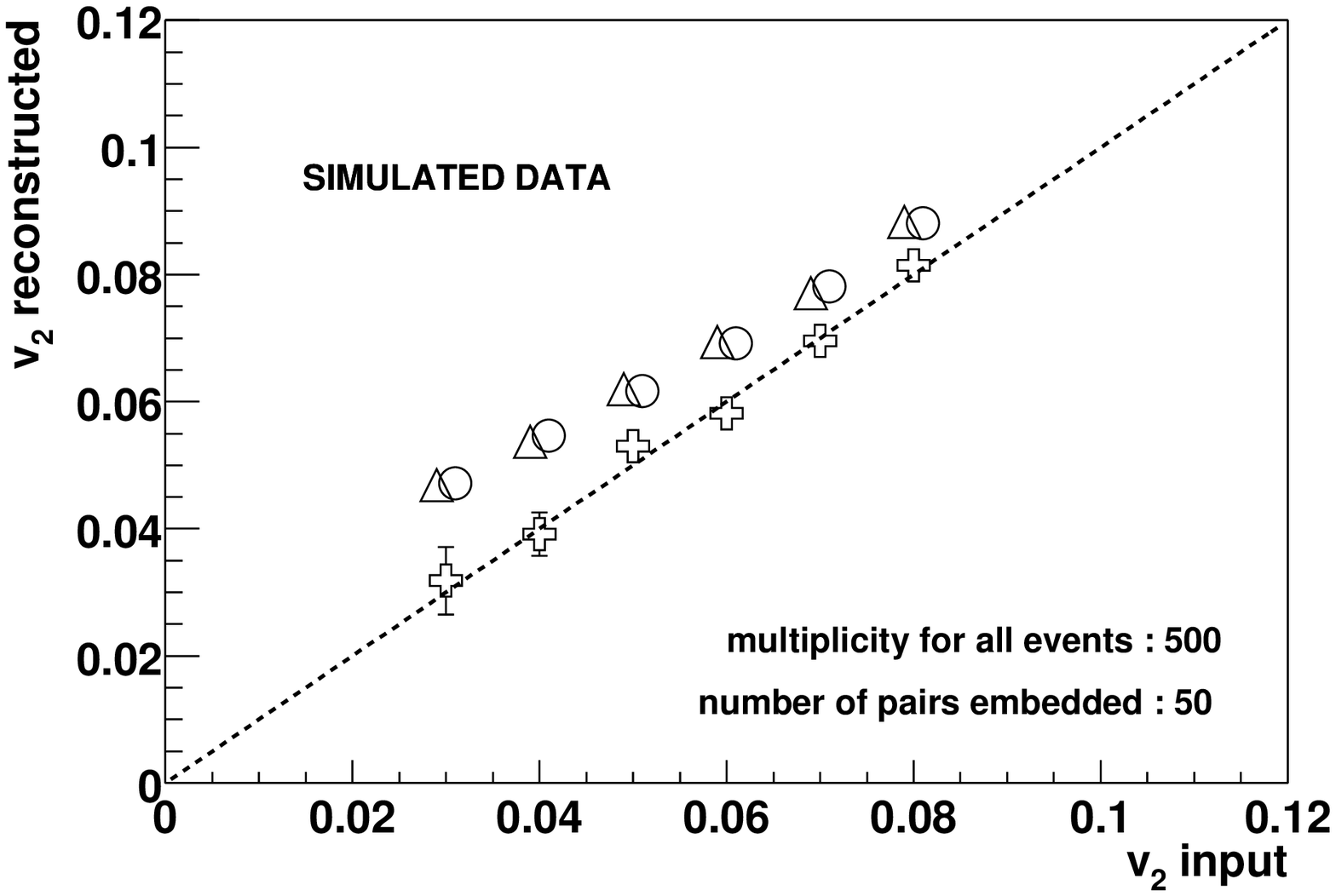}}
\caption{ Reconstructed $v_2$ versus input $v_2$, with a dashed line to mark 
the ``input = output'' diagonal.  The circles, triangles and crosses represent 
$v_2$ from the conventional method, from the 2nd-order cumulant method, and 
from the 4th-order cumulant method, respectively. }

\end{figure}

\section{Conclusion}

It is concluded that quadruplet correlation analyses can reliably separate 
flow and non-flow correlation signals.  The cumulant approach applied to 
the year-one STAR data reported in this work has demonstrated some advantages 
over the previous alternative approaches for treating non-flow effects.  In 
particular, the cumulant approach is sufficiently flexible that we have now 
chosen to present $v_2$ measurements corrected for non-flow effects, in 
contrast to the earlier analyses where the non-flow contribution was partly 
removed and partly quantified by the reported systematic uncertainties.  
On the other hand, a 4th-order cumulant analysis is subject to larger 
statistical errors than a conventional pair correlation analysis of the same
data set.  In the case of year-one data from STAR, the intrinsic advantages 
of a higher-order analysis are offset by the increased statistical 
errors, but in the case of the forthcoming higher statistics running in 2001
 and beyond, a higher-order analysis will provide a clear advantage.  

It is observed that non-flow correlations are present in $\sqrt{s_{NN}} = 
130$ GeV Au + Au events throughout the studied region $|\eta| < 1.3$ and 
$0.1 < p_t < 2.0$ GeV$/c$, and are present at all centralities.  The 
largest contribution from non-flow correlations is found among 
peripheral collisions.

\begin{theacknowledgments}
We thank Nicolas Borghini and Jean-Yves Ollitrault for helpful discussions and advice. 
\end{theacknowledgments}

\end{document}